\documentclass[runningheads]{llncs}
\usepackage[T1]{fontenc}
\usepackage{graphicx}
\usepackage{graphicx}%
\usepackage{multirow}%
\usepackage{hyperref}%
\usepackage{amsmath,amssymb,amsfonts}%
\usepackage{mathrsfs}%
\usepackage[title]{appendix}%
\usepackage{xcolor}%
\usepackage{textcomp}%
\usepackage{manyfoot}%
\usepackage{booktabs}%
\usepackage{algorithm}%
\usepackage{algorithmicx}%
\usepackage{algpseudocode}%
\usepackage{listings}%

\usepackage{longtable}%
\usepackage{changepage} 
\usepackage{setspace}   
\usepackage{ragged2e}   
\usepackage{comment}

\usepackage{tabularx} 
\usepackage{lscape}
\usepackage{adjustbox}
\usepackage{caption}

\usepackage{background}
\backgroundsetup{
  position=current page.south east,
  angle=90,
  nodeanchor=west,
  vshift=30mm, 
  hshift=80mm,  
  color=gray,
  opacity=1,
  scale=0.8,
  contents={\parbox{20cm}{\raggedright\footnotesize
  Submitted to CAiSE 2025.
  This version of the contribution has been accepted for publication, after peer review but is not the Version of Record and does not reflect post-acceptance improvements, or any corrections. The Version of Record is available online at: \url{https://doi.org/[insert DOI]}. Use of this Accepted Version is subject to the publisher's Accepted Manuscript terms of use \url{https://www.springernature.com/gp/open-research/policies/accepted-manuscript-terms}.
  }}
}

\begin{document}
\title{Evaluating Organization Security: User Stories of European Union NIS2 Directive}
\titlerunning{Evaluating Organization Security: USs of NIS2}
%

\author{Mari Seeba\inst{1, 2}\orcidID{0000-0002-9066-2467} \and
Magnus Valgre\inst{1} \orcidID{0009-0000-2575-5301}  \and
Raimundas Matulevičius\inst{1}\orcidID{0000-0002-1829-4794}
}
\authorrunning{M. Seeba et al.}
%
\institute{Institute of Computer Science, University of Tartu, Estonia \\
\and
Estonian Information System Authority\\
\email \{mari.seeba, magnus.valgre, raimundas.matulevicius\}@ut.ee}
%
\maketitle              
\begin{abstract}

The NIS2 directive requires EU Member States to ensure a consistently high level of cybersecurity by setting risk-management measures for essential and important entities.
Evaluations are necessary to assess whether the required security level is met. This involves understanding the needs and goals of different personas defined by NIS2, who benefit from evaluation results. In this paper, we consider how NIS2 user stories support the evaluation of the level of information security in organizations.
Using requirements elicitation principles, we extracted the legal requirements from NIS2 from our narrowed scope, identified six key personas and their goals, formulated user stories based on the gathered information, and validated the usability and relevance of the user stories with security evaluation instruments or methods we found from the literature.
The defined user stories help to adjust existing instruments and methods of assessing the security level to comply with NIS2. On the other hand, user stories enable us to see the patterns related to security evaluation when developing new NIS2-compliant security evaluation methods to optimize the administrative burden of entities.

\end{abstract}
\section{Introduction}\label{sec:Intro}


In 2015, the enactment of the European (EU) GDPR data protection regulation changed the attitude toward data privacy and raised awareness of the issue \cite{szczepaniuk2020information}. The impact of the implementation of the GDPR has been global. Similar data protection regulations have now been established all over the world \cite{WorldPrivacyForum}.
With the NIS2 Directive, the aim of the EU Commission \cite{NIS2} is to change the information security management postures of organizations in the EU to effectively protect the digital single market, and reduce the damaging impacts of security incidents on the economy and society \cite{NIS2}. Similarly to the enactment of GDPR, a widespread increase in security awareness and implementation of the directive's requirements outside the EU is expected.

From a policymaker's perspective, NIS2 creates explicit measures for entities required to implement the directive's requirements \cite{VANDEZANDE2024105890,Grigaliūnas2024}. From the perspective of implementers and engineers, the complexity of interpreting and implementing the regulations is recognized
\cite{Islam2010255,Kiyavitskaya2008,fatema2016}.
For engineers, security is relative, depending on many factors, and the \textit{all-hazards approach} used in NIS2 \cite{NIS2} is an unattainable situation. Therefore, it is appropriate to reformulate the requirements of the regulations into a format understandable to engineers. This allows policymakers and implementers to break out of their silos and get involved in a dialogue, and exchange feedback on the practical effectiveness of the directive \cite{pisa2020governing,alsaadi2019}. Additionally, implementing the regulations should not involve reinventing the wheel but rather building on existing standards and solutions for harmonization \cite{tasheva2022}.

In this paper, we focus on one of the NIS2 directive's objectives - achieving a common high level of security across the EU \cite{NIS2}. 
We narrowed our research area to the evaluation of the implementation level of risk-management measures, which are presented in NIS2 to organizations. 
We analyzed the regulation to identify who must meet the specified requirements and for what purpose, based on data from the assessment of organizations' information security levels.
Ultimately, we elicit user stories related to the results of organizations' information security level evaluation support that meet NIS2 requirements.
Using the requirement engineering method, we discovered 6 personas and 10 user stories that are directly related to an organization's security level evaluation results. Identified user stories are the prerequisite for identifying ways for NIS2 implementation.


\section{Backgound}\label{sec:BG}
\subsection{NIS2}
The NIS2 Directive \cite{NIS2} was published in the Official Journal of the EU and entered into force on January 16, 2023. It aims to establish a high level of cybersecurity across the Union to protect the European single market from security incidents that could disrupt the economy and society \cite{NIS2}. 
NIS2 provides risk-management measures for entities, addresses communication channels and reporting, defines contact points during security incidents, and guides supervisory activities and penalties. 
The organizations who are required to comply with NIS2 are public or private 
entities in the high-
criticality sectors, such as energy, transport, 
drinking and wastewater, public administration, digital infrastructure, and others listed in Annexes I and II of NIS2 \cite{NIS2}. 
Compared to NIS1 (published 2016) the NIS2 added more than 10 obliged sectors. 
Therefore, the requirements must be explicit to the implementing entities and achievable with reasonable investment and administrative costs. EU Member States (MS) had to transpose NIS2 into local law by the 17th of October, 2024 \cite{NIS2}. 

Eliciting requirements from a legal text is complicated because (i) the text is fragmented and uses concepts and terminology that are different from software engineering, (ii) requirements can arise from different levels of law (e.g., EU level or the member state level or regulative standard), (iii) imperfection and vagueness of the law and its wording allows multiple interpretations and (iv) dynamics of the law over time \cite{Kiyavitskaya2008,Islam2010255}. 
To mitigate risks (ii) and (iv), we only considered the EU-level NIS2 directive, which all Member States must adopt based on the principle of minimum harmonization stated in Article 5 \cite{NIS2}. That means the Member State must adopt NIS2 as the minimum baseline. We do not address the level of Member State's requirements. We only used the version of NIS2  \cite{NIS2} from 2022. To mitigate risks (i) and (iii), we used methods described in Sec~\ref{sec:Method}.

\subsection{Security evaluation}
Various methods and instruments \cite{khaleghi2022comprehensive,leszczyna2021review,rea-guaman_maturity_2017} can be used to assess an entity's security level, which can be done through direct measurements of risk management measures, self-assessments (e.g., security maturity models), or second or third-party evaluations (e.g., audits, penetration testing). More indirect measurements can also be used to assess the level of security, such as counting the number of organizations that hold some kind of security management certificate (e.g., ISO/IEC27001 Information security management system compliance certificates). This study explores how the security level evaluation results of NIS2-obliged entities can be interpreted and applied to evaluate risk-management measures implementation independently of any specific evaluation method or instrument.

\section{Related Work}\label{sec:RW}
Only six of the 27 Member States (Belgium, Croatia, Hungary, Italy, Latvia, and Lithuania)\footnote{\url{https://dnsrf.org/nis2-transition/} NIS2 transition tracker status by 2024-11-01} succeeded in transposing the NIS2 into local law by October 17, 2024. One of the reasons for the delay in this transposition is the different views of lawyers and implementers on the applicability of the legal text.

 Juridical publications (e.g., \cite{VANDEZANDE2024105890,CHIARA2024105961}) see the NIS2 Directive primarily as an enabler of raising the security levels of the Member States and emphasize sanctions for non-compliance, and describes NIS2 as a regulation with explicit requirements. In the view of the engineers, the intricate structure and complex legal language of the texts cause questions and ambiguities \cite{hassani2024rethinking,Grigaliūnas2024}. From the legal point of view, the primary concern is about the excessive administrative burden \cite{VANDEZANDE2024105890}. However, this originates from the separate implementation of every single clause rather than a comprehensive information security management system, which would align with best practices (e.g., the international standard ISO/IEC 27001 controls \cite{ISO27001}) and support the entities' whole management system.
If the lawyers recommend balancing requirements in implementing regulations \cite{VANDEZANDE2024105890}, legal text analysis by engineers would instead find optimization patterns and holistic models \cite{hassani2024rethinking,Jorshari2012,Grigaliūnas2024}.

There are few references to the analysis of evaluating the security level concerning NIS2.  Wanecki et al. \cite{Wanecki2023} developed a cybersecurity model based on NIS2 but did not cover the evaluation of the achieved security level. The only option mentioned is conducting audits, which only cover essential entities. Grigaliunas et al. \cite{Grigaliūnas2024} created a GDPR, NIS2, and ISO/IEC 27001-based framework that categorizes controls into preventive, detective, and corrective; allowing entities to align their security maturity levels but without considering other stakeholders' expectations on security level evaluations.

Fatema et al. \cite{fatema2016} first extracted the relevant legal clauses and eliminated the irrelevant to determine the relevant scope.
Hassani et al. \cite{hassani2024rethinking}, using LLM-s for legal compliance analysis, turned attention to legal text sentences that are not separate units. It is essential to follow the sequence, definitions, and cross-references as a whole. It is not an option to treat individual sentences out of context. Here the personas and their relationship models can be helpful. Therefore, the legal text analysis cannot start with extracting relevant information, the entire text of the regulation must first processed.

Legal text ambiguity patterns (lexical, analytical, vagueness, and generality) are described by Alsaadi et al. \cite{alsaadi2019}, who studied the EU Medical Device Regulation (MDR) and the Health Insurance Portability and Accountability Act (HIPAA). After analyzing the text, they set a goal to remove the ambiguity by rephrasing the legal requirements into user stories. They used more relevant terms to make user stories unambiguous for engineers and enable discussions about the personas' activities and their purposes. 

Based on the previously described references, we performed our requirements elicitation. The method is described in more detail in the next section.

\section{Creating User Stories}
\subsection{Method}\label{sec:Method}

We elicited the personas, their goals and dependencies, and user stories. As our study is based on the NIS2 Directive's \cite{NIS2} legal text, we also followed the suggestions on legal text analysis described by \cite{Jorshari2012,fatema2016,Islam2010255}. 

Following the example of \cite{fatema2016}, after reading the entire NIS2 directive text, we only selected the clauses relevant to our study. 
Then we analyzed the sentences individually, marking actors, actions, and resources as suggested by Islam et al. \cite{Islam2010255}. 
Following the steps outlined in \cite{Jorshari2012,Islam2010255}, we 
created the strategic dependency goal model of the personas in 
i* modelling language \cite{yu2010social}.

Next, we defined the user stories (see in Sec.~\ref{sec:UserStoriesDescription}) related to the organization's security level evaluation following the template format proposed by Cohn \cite{Cohn2004US,Cohn2017US-s}:

\begin{verbatim} 
    As a <type of user>, I can <some goal> so that <some reason>.
\end{verbatim}
The simple template-based structure is understandable to stakeholders and software engineers by helping to reach a common understanding of the requirements and define the quality guideline \cite{Lucassen2016USusability}. 
To validate the user stories, we aligned them with existing methods and instruments to demonstrate that the use cases they covered already exist in practice.


\subsection{NIS2 requirements elicitation}\label{sec:NIS2req}

Our scope is to find from NIS2 the clauses related to organizations' (in NIS2 vocabulary - essential and important entities) security level evaluation. At first, we got acquainted with the whole NIS2 text, and we highlighted the relevant clauses that can be interpreted in the context of security evaluation of entities. 
We also used searches for keywords such as \textit{ensure}, \textit{level}, \textit{assess*}, \textit{oversee}, and \textit{measures} for crosschecking. 
Keywords were chosen based on interpretation options:
at the Member State level, the term \textit{ensure} can be interpreted as requiring the Member State to evaluate and measure entities’ security levels. This evaluation process ensures that entities comply with regulations and maintain an expected level of cybersecurity. To get the \textit{oversee}, an activity related to evaluation is needed. A term \textit{level} describes an association with something, which refers to measurement, assessment, or evaluation. We selected string \textit{measure*} to find the relationship between risk-management measures, as well as the relationship between measurement.
In the Appendix \ref{appendix:NIS2clauses} are shown the filtered clauses, which relate to entities and their security evaluation. 

To identify the personas and find their dependencies and goals, we analyzed all selected clauses of NIS2 using \cite{Islam2010255} legal text analysis model steps. In the legal text, we marked personas or subjects of action as \underline{underlined}, normative phrases and modal verbs are marked in \textbf{bold}, and actions in \textit{italics} like:

Art20(1) ``\underline{Member States} \textbf{shall} \textbf{ensure} that the \textbf{\underline{management bodies of}} \textbf{\underline{essential and important entities}} \textit{approve the cybersecurity} \textit{risk-management measures} \textit{taken} by those \underline{entities} in order \textit{to comply with Article 21}, \textbf{oversee its implementation} and can be held liable for infringements by the entities of that Article.'' \cite{NIS2}

This allowed us to pick out the preliminary mentioned personas (actors): ENISA, European Parliament, peer reviewers, Member State, small and medium-sized enterprises, management bodies of essential and important entities, important entities, essential entities, service providers \& suppliers, and competent authority for supervisory. 

We excluded the European Parliament from this list as it is outside the scope of organization-level security evaluation. Also, we excluded the organization's internal structure and processes and focused only on the organization as a general entity with its management body and employees. We engaged peer reviews under the Member State persona, as the process of peer reviews is organized at the Member State level in cooperation with cybersecurity experts from at least two Member States. 
In our security evaluation scope, the essential and important entities differ only in the supervisory context, where essential entities should be subject to a comprehensive supervisory regime (preventative and after security incidents). In contrast, important entities should be subject to a simplified supervisory regime after a security event or someone's hint of an entity security violation. Security level evaluation is similar in both cases. We also included small and medium-sized enterprises under the persona Entity and Suppliers or Service Providers because all Entities can simultaneously be someone's Service Provider \& Supplier and essential and important entities.

Additionally to already mentioned personas,  from NIS2 recitals No 56, we found the persona called \textit{Member State point of contact} for small and medium-sized enterprises, who should guide and assist small and medium-sized enterprises regarding cybersecurity-related issues. Impersonally, but the same guidance and assistance issue is mentioned as an expected clause of Member State National cybersecurity policy (Art 7(2)(f) and (i)~\cite{NIS2}). To avoid confusion with another single point of contact used for different processes described in NIS2, we named this guidance and assistance provider as a Security Consultant. 

So, we limited the personas of NIS2, who are relevant in the context of organization security level evaluation with the list:  \textit{Member State}, \textit{Supervisory Authority}, \textit{ENISA}, \textit{Entity}, \textit{Security Consultant} and \textit{Service Provider \& Supplier}.
In the next section, based on legal text analysis, we describe the personas mutual relations goal model.

\subsection{Personas' Dependency Model}\label{sec:NIS2Personas}

The six personas rely on organizations' security data or generalized results to achieve their objectives. The goal of the \textbf{Member State} is to receive secure services from Entities, obtain the service's security statuses, assign Consultants to support Entities with cyber-security issues, assign a Supervisory Authority and provide guidance and training on cybersecurity to Entities (\cite{NIS2} Articles: Art1(1); Art7(2); Art19(1.a); Art20(1),(2); Art21(1),(2),(3),(4); Art31(2), Art32(2),(4); Art33(2)). 
\textbf{Supervisory Authority} is assigned by Member State. It should provide feedback on Entities' security status (\cite{NIS2} Articles: Art21(1),(2),(4); Art31(2); Art32(2),(4); Art33(2)). \textbf{ENISAs'} goal is to evaluate the security status of Member States and Entities and provide results to the EU Parliament so that it could assess the EU security level (\cite{NIS2} Articles: Art1(1), Art18(1)). 
\textbf{Security Consultants} assist and guide Entities on risk-management measures implementation and could be assigned by Member State (\cite{NIS2} Articles: Art 7(2); Art20(2); Art21(2)). \textbf{Entity} provides secure services to Member State and follows Member State regulations (implements risk-management measures, passes training). It also gets secure services and products from Service Provider \& Supplier (\cite{NIS2} Articles: Art7(2); Art20(1),(2); Art21(1),(2),(3),(4); Art32(2),(4); Art33(2)). \textbf{Service Provider \& Supplier provides} provide secure services or products to Entity, (\cite{NIS2} Articles: Art21(2); Art21(3)). 

\begin{figure}
    \centering    \includegraphics[origin=c,width=1\linewidth]{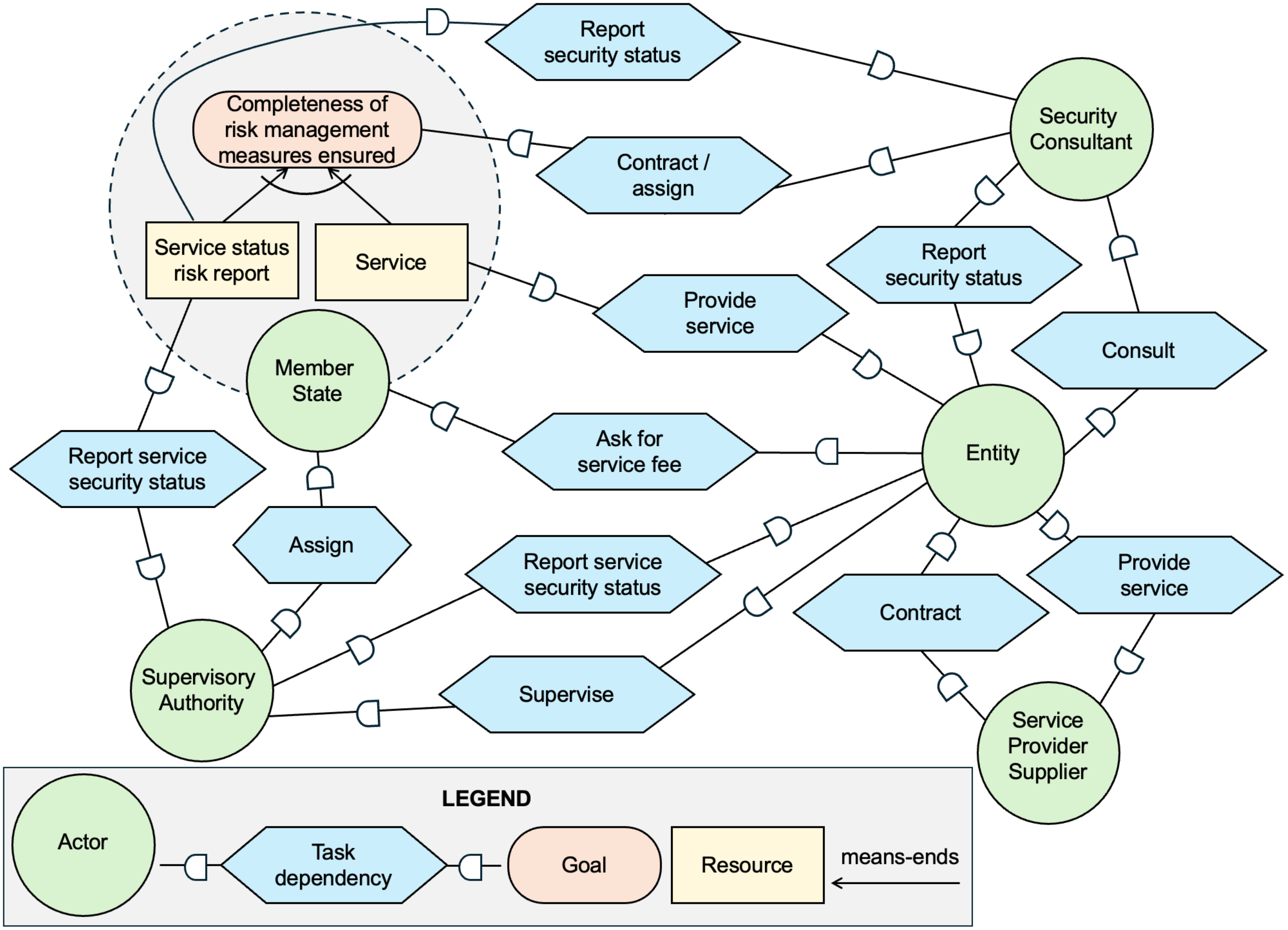} 
    \caption{Personas' Dependency Model} 
    \label{fig:Goals}
\end{figure}

We illustrate the above dependencies in Fig.~\ref{fig:Goals}. The model emphasizes the personas' dependencies and supports the user stories. For simplification, ENISA is not included. However, as described above, ENISA obtains the security status of the Member State and Entities and shares the best practices with other Member States. The prioritization and optimization of activities is the task of the Supervisory Authority. 
It should be noted that a specific organization can take different roles. For instance, an Entity can simultaneously be a Service Provider, Supplier, and Security Consultant. In some cases, the Entity can be a Member State or Supervisory Authority (e.g., Computer Security Incident Response Team, CSIRT).

\subsection{Security Evaluation User Stories}\label{sec:UserStoriesDescription}

Next, we describe the User Stories. They are formulated based on the legal requirements quoted in the Appendix~\ref{appendix:NIS2clauses}) 
and the dependencies illustrated in Fig.~\ref{fig:Goals}. User Stories are divided into groups based on Personas. The result is presented in Table~\ref{tab:US}, where the 
goals regarding security level evaluations are described for each Persona. We also included references to NIS2 clauses. 

\begin{longtable}[c]{|p{1.5cm}| p{10.4cm}|} 
    \caption{User Stories of NIS2 \cite{NIS2} Related to Security Level Evaluation of Entities}
    \label{tab:US}
    \small
    
 \endfirsthead

    \hline
    
 \multicolumn{2}{|c|}{\textit{Continuation of Table \ref{tab:US}}}\\
 \hline
 \hline
 \endhead
 \endfoot
        \hline
        Role: &  \textbf{Member State} \\
        \hline 
        Goal: & \small Factual proof of achieving a high common level of cybersecurity in all sectors and entities to avoid cyber incidents causing major damage to economics and society. \\
        Reference: & \small Art1(1); Art7(2); Art19(1.a); Art20(1),(2); Art21(1),(2),(3),(4) of NIS2~\cite{NIS2} \\
        \hline
        \multicolumn{2}{|p{11.9cm}|}{\textbf{US1.1:} \small As a Member State, I can oversee the security posture of Entities through structured security level evaluation results, so that I achieve awareness of compliance with regulations.}
        \\
        \multicolumn{2}{|p{11.9cm}|}{\textbf{US1.2:} \small As a Member State, I can evaluate an entity’s cybersecurity level using an all-hazards approach, so that I can allocate resources to address directly on identified vulnerabilities.} \\
\hline
\hline
         Role:&  \textbf{Supervisory Authority}\\
         \hline
         Goal:& \small The Supervisory Authority should base on risk assessments when planning their supervisory tasks, but they should optimize the workflow and not unnecessarily hamper the business activities of the entity concerned. \\
         Reference:&  \small Art21(1),(2),(4); Art31(2); Art32(2); Art32(4); Art33(2)of NIS2~\cite{NIS2}\\
        \hline
        \multicolumn{2}{|p{11.9cm}|}{\textbf{US2.1} \small As a Supervisory Authority, I can prioritize supervisory tasks by using all hazard-covering security evaluation results so that I can focus supervisory tasks on high-risk entities or areas.}\\
        \multicolumn{2}{|p{11.9cm}|}{\textbf{US2.2} \small As a Supervisory Authority, I can ensure (with a security evaluation instrument) that entities that did not comply with regulatory requirements implement corrective risk-management measures within reasonable deadlines so that supervisory resources are used effectively and unnecessarily hamper the business activities of the entity is avoided.}\\
\hline
\hline
        Role: &  \textbf{ENISA} \\
        \hline
        Goal: & \small Collect data to evaluate EU security level and report the result to EU Parliament \\
        Reference: & \small  Art1(1), Art18(1) of NIS2~\cite{NIS2} \\
        \hline
        \multicolumn{2}{|p{11.9cm}|}{\textbf{US3.1} \small As ENISA, I can collaborate with Member States to assess collected evaluation data on cybersecurity capabilities and awareness, so that I can share cybersecurity best practices and gaps across the European Union.}\\
\hline
\hline
        Role: &  \textbf{Security Consultant} \\
        \hline
        Goal: & \small Consultants should help improve the entity's security level by finding and focusing on vulnerable areas of the entity. \\
        Reference: & \small  Art20(2); Art21(2) of NIS2~\cite{NIS2} \\
        \hline
        \multicolumn{2}{|p{11.9cm}|}{\textbf{US4.1} \small As a security consultant, I can get an overview of the entity's security maturity evaluation results so that the most vulnerable areas can be prioritized in a timely manner for an improvement plan.}
        \\
        \multicolumn{2}{|p{11.9cm}|}{\textbf{US4.2} \small As a security consultant, I can re-evaluate the entity's risk-management measures implementation so that tracking characterizes risk-management measures implementation status progress.} \\
        
\hline
\hline
        Role: &  \textbf{Entity} \\
        \hline
        Goal: & \small To obtain an overview of the entity's cybersecurity risk-management measures all-hazard approach to confirm security status and improve vulnerable areas  \\
        Reference: & \small Art20(2); Art21(2) of NIS2~\cite{NIS2} \\
        \hline
        \multicolumn{2}{|p{11.9cm}|}{\textbf{US5.1} \small As an entity, I can ensure the entity adopts an all-hazards approach\footnote{Some recognized standard like ISO27001\cite{ISO27001} can limit the uncertainty of the all-hazard approach.} to cybersecurity so that the evaluation results show strengths and direct to plan improvements to our security shortcomings.} \\
\hline
\hline
        Role: &  \textbf{Service Provider \&  Supplier} \\
        \hline
        Goal: & \small Get an evaluation of Service Provider \&  Supplier security to share with partners and assess compliance with partner requirements \\
        Reference: & \small Art21(2); Art21(3) of NIS2~\cite{NIS2} \\
        \hline
        \multicolumn{2}{|p{11.9cm}|}{\textbf{US6.1} \small As a Service Provider \&  Supplier, I can provide the risk-management measures in all-hazard evaluation approach results to the partner Entity so that the Entity can choose us as the most suitable secure suppliers.}
        \\
        \multicolumn{2}{|p{11.9cm}|}{\textbf{US6.2} \small  As a Service Provider \&  Supplier, I can regularly evaluate my cybersecurity practices so that I can present my evaluation results to my partner Entity to demonstrate our security.} \\
        \hline
        \hline
\end{longtable}

To explain the user stories provided in the Table~\ref{tab:US}, we will take a closer look at how to use user stories and also show  how user stories are connected and need additional analysis by the Member State (MS).

US2.1 calls for the Supervisory Authority to prioritize activities in the context of all risks and plan its activities based on this. In essence, this prepares a sample of organizations to focus on. MSs can have several security-related supervisory authorities with different focuses in terms of sectors and functions (e.g., in Estonia, there are three: NCSC-EE, whose supervision deals with cybersecurity in general; the Financial Supervision Authority focuses on the financial sector, and the Data Protection Authority monitors data protection. In Finland, oversight of cybersecurity issues is divided between 8 institutions by area.) Each supervisory unit must prepare its plans based on its objectives and, in addition, national risk assessments.

Therefore, a Supervisory Authority needs aggregated data collected from organizations (reusing the data collected during US5.1) that distinguishes between relevant sectors and their vulnerabilities. This focuses the monitoring process on identifying causes and finding inputs for improvement according to sectors (e.g., finance, energy, research) or security functions (e.g., data protection, incident management, cyber hygiene, and awareness). 

However, the MS must specify which metadata must be collected during the US5.1 process for filling in the US2.1 (but also US1.1, US1.2, US3.1). This could depend on MS supervisory authorities, their sectoral affiliation, and how detailed the aggregation needs to be. 
The questionnaire or instrument detail level should match the supervisory focus needs and  data protection requirements, balancing aggregation and listing  specific technical measures.

\subsection{Validity of User Stories}\label{sec:Validation}

We validated the user stories by aligning them to existing or proposed security evaluation instruments. The instruments were chosen to cover a wide spectrum of applications and to show that the user stories described in this paper and in NIS2 reflect the situation in the real world.
The reviewed instruments can be divided into the following categories: publications \cite{Seeba2022F4SLE,bernik2016measuring,prislan2020real,maleh2017towards,rae2019defining,malaivongs2022cybertrust,jazri2018measuring,you2016advanced}, 
ENISA or state-sponsored tools \cite{KybermittariWeb,ENISAAssessmentSMEs,NUKIBReport,irelandSelfAssessment,IASMEEssentials,spainSelfAssessment,greeceSelfAssessment},
cybersecurity indices \cite{EUCSIWeb,NCSIWeb,GCIWeb,BelferCent2020NCPI,CDIWeb}, 
maturity models \cite{C2M2Tool}, 
and official audits \cite{EstoniaAuditLocal,EstoniaAuditXroad,latviaAudit2022eservices}.

\begin{table}[]
\centering
\caption{Instruments that implement user stories: '+' instrument covers the given user story; '-' instrument does not cover the user story; 'v' instrument can be used to cover the user story, but it is not explicitly meant for that purpose; '*' instrument is not usable \textit{as-is} and needs significant effort to be workable.}
\label{tab:tools}
\small
\begin{tabular}{@{}|m{0.6\textwidth}|l|l|l|l|l|l|@{}}
\hline
\textbf{Instruments}                                                                                                                                                                                   & \textbf{US1} & \textbf{US2} & \textbf{US3} & \textbf{US4} & \textbf{US5} & \textbf{US6} \\ \hline
F4SLE \cite{Seeba2022F4SLE},  Kybermittari \cite{KybermittariWeb}, Jazri et al. \cite{jazri2018measuring}*                                & +   & +   & v   & +   & +   & +   \\ \hline
Bernik et al. \cite{bernik2016measuring}*, Prislan et al. \cite{prislan2020real}*, Maleh et al.\cite{maleh2017towards}*, Rae and Patel \cite{rae2019defining}*, Malaivongs et al. \cite{malaivongs2022cybertrust}*, You et al. \cite{you2016advanced}*, self-assessment tools by Ireland \cite{irelandSelfAssessment} and Greece \cite{greeceSelfAssessment}, C2M2 Maturity Model \cite{C2M2Tool}  & -   & -   & v   & v   & +   & v   \\ \hline
NÚKIB Report\cite{NUKIBReport}*                                                                            & +   & +   & -   & -   & -   & -   \\ \hline
EU CSI \cite{EUCSIWeb}*                                                                     & +   & -   & +   & -   & -   & -   \\ \hline
Cybersecurity Indices NCSI \cite{NCSIWeb}*, GCI \cite{GCIWeb}*, NCPI \cite{BelferCent2020NCPI}*, CDI \cite{CDIWeb}*,               & +   & -   & -   & -   & -   & -   \\ \hline
Official Audits by Estonia \cite{EstoniaAuditLocal}*, \cite{EstoniaAuditXroad}* and Latvia \cite{latviaAudit2022eservices}*          & +   & v   & -   & -   & -   & -   \\ \hline
Self-Assessment tools by ENISA \cite{ENISAAssessmentSMEs}, IASME \cite{IASMEEssentials} and Spain \cite{spainSelfAssessment}        & -   & -   & -   & v   & +   & v   \\ \hline
\end{tabular}
\end{table}

An overview of the instruments and their coverage of user stories is illustrated in Table~\ref{tab:tools}. Each covers at least one of the user stories. The instruments have been created for different purposes and levels of abstraction.  Some instruments compare and describe the cybersecurity postures of countries on a global scale (e.g. \cite{EUCSIWeb,NCSIWeb,BelferCent2020NCPI,CDIWeb}), while other instruments are meant for individual organizations (e.g. \cite{spainSelfAssessment,greeceSelfAssessment,irelandSelfAssessment}) and comply with the Entity, Security Consultant or Service Provider \& Supplier user stories.

Different abstraction levels require different approaches, which can lead to loss of detail. 
Cybersecurity indices \cite{EUCSIWeb,NCSIWeb,BelferCent2020NCPI,CDIWeb} compare the security postures of entire countries. However, they do not consider the differences in the levels of digitalization, that determines the actual required level of security. Indices rely on high-level data, such as the existence of appropriate legislation and cybersecurity-related institutions, but these facts will not be helpful for individual entities in determining their security level. Still, indices can contain some data on entities (e.g., how many organizations have attained some specific security certifications \cite{EUCSIWeb}). The EU Cybersecurity Index \cite{EUCSIWeb} created by ENISA uses data gathered from EU member states and is also the subject of US3.

From the bottom-up perspective, most of the methods available are created to help individual organizations perform a self-assessment to find areas of improvement (e.g. \cite{spainSelfAssessment,IASMEEssentials}) and they lack functionality in aggregating data and presenting it at a higher level of abstraction. Still, they fulfill the goals of US5 but can also cover US4 and US6.

The least covered is US2, which means the needs of governmental overseeing bodies are the least considered by the currently available instruments. Two notable instruments here are the audits performed by national audit offices \cite{EstoniaAuditLocal,latviaAudit2022eservices} and the annual report \cite{NUKIBReport} composed by the National Cyber and Information Security Agency of the Czech Republic (NÚKIB). Due to their thorough nature, audits bring a lot of insight into a given topic and provide concrete recommendations for improvement, but they are not periodically done on the same topic and this limits their usefulness in verifying that improvements were implemented. 
The NÚKIB report \cite{NUKIBReport} gathers its data surveys of entities without any individual feedback to the Entity. 

There is a real lack of instruments and methods that help collect security evaluation data on individual entities and bring it together for central decision-makers but also, at the same time, provide the given organization with feedback on their current capabilities and areas that need enhancement. In other words, there are only a few tools \cite{KybermittariWeb,Seeba2022F4SLE} that simultaneously cover the User Stories related to Entities and MSs.

Few security evaluation instruments directly corresponding to NIS2 requirements have yet been created \cite{Seeba2022F4SLE,KybermittariWeb}. However, an ENISA security assessment pilot report based on NIS2 Article 18 \cite{NIS2} has already been completed, but only for internal use and not publicly available. 

The analysis showed that the tools tend to be mono-functional, but some are multifunctional (e.g. \cite{bernik2016measuring,irelandSelfAssessment}), simultaneously covering several user stories. Instruments are divided into Member State (US1-US3) and Entity goals (US4-US6). 
From Table~\ref{tab:tools}, we can see that all our user stories are covered at least by one instrument, showing that user stories are realistic. However, not all instruments are usable for each user story or suitable for periodic/repeatable evaluation.

\section{Discussion}\label{sec:discussion}
A security evaluation aims to provide situational awareness and present the dynamics of security. During the validation of the user stories, we identified different security evaluation methods and instruments which fulfill the needs of either the Member States or the Entities. 


Security risk management is an iterative process. If an adversary is able to identify a single vulnerability, they are able to damage the system. However, the Entity (i.e., defender) must implement multiple security countermeasures to mitigate various security risks. This task requires considerable resources. Any activity (including an evaluation of the security level) that does not directly contribute to risk mitigation can significantly burden the organization's administration. Therefore, evaluating the security level should not be a goal in itself; rather, it must be an integrated part of security management and create value for the Entity. We observed four personas (Member State, Supervisory Authority, Consultant and ENISA) who require security evaluation input from entities to support their tasks. The needs of other stakeholders should be integrated into the Entity's security evaluation process. The user stories could help achieve this goal of finding ways to reuse data, optimize, automate, and manage security evaluation, especially from the Entity's point of view.\\
Finding patterns within legal requirements is the job of experts and engineers. So is optimizing or balancing burdensome activities, as described in Sec.~\ref{sec:RW} by \cite{hassani2024rethinking,Jorshari2012,Grigaliūnas2024}. Our research showed that the NIS2 analysis allowed us to find optimization points that could simplify the implementation of NIS2 in a less burdensome way.

\noindent\textbf{Limitation:} Our scope is narrowed to NIS2 and does not expand to other EU regulations.
User stories follow the NIS2 Directive \cite{NIS2} and are written at a high level of abstraction. This approach allows us to avoid conflicts with the local laws of the Member States when the Member States have transposed NIS2. It also ensures the instantiation of user stories in Member States by adding contextual details.
The flexibility of the user stories may lead to challenges in implementation, as they are not explicitly aligned with specific standards or tools.

NIS2 does not refer to any reference standards, the user stories must remain flexible in their application. However, this flexibility may limit the coverage of user stories and their scope in the security assessment. For example, information security management standards (e.g., ISO/IEC27002 \cite{ISO27002:2022}) cover the risk-management measures detailed in Article 21(2) of NIS2, but these standards do not specify NIS2-compliant reporting during incidents. Similar concerns are observed for the security awareness training and awareness evaluation of the management boards. However, the Estonian Information Security Standard (E-ITS) \cite{E-ITS}, a detailed catalog of measures and guidelines, provides corresponding instructions. 

The user stories do not explicitly address the continuous compliance required by NIS2, as this is ensured by default through repeated security level evaluations.


\section{Concluding Remarks}

In this paper, we identified six stakeholders (personas) from the NIS2 directive who depend on the results of an organization's (entity's) security level evaluations to fulfill their tasks. 
We created user stories that reflect each persona's relation to the evaluation results. The user stories are not dependent on any standards or instruments used for security evaluations. Instead, the user stories are described at the general level.

When adopting NIS2 and planning the security evaluation activities, a Member State should consider how to avoid overloading organizations. Different personas might behave in ways that are based only on their individual needs. The defined user stories could support the planning process by reusing security evaluation results without overburdening the organizations.
\paragraph{Further Work}
As we are developing the FASLE instrument \cite{Seeba2022F4SLE,Seeba2023MASS_F4SLE,Seeba2023MUSE}, we work with stakeholders to test all the described user stories with F4SLE in real-world situations to achieve their operational objectives. This way, we can detail the user stories at the national level to show how they can be implemented with the instrument that collects data only once and uses it for different stakeholders.

\begin{credits}
\subsubsection{\ackname} This work is part of the Cyber-security Excellence Hub in Estonia and South Moravia (CHESS) project funded by the European Union under Grant Agreement No. 101087529. 
Views and opinions expressed are however those of the author(s) only and do not necessarily reflect those of the European Union or European Research Executive Agency. Neither the European Union nor the granting authority can be held responsible for them.

\subsubsection{\discintname}
The authors have no competing interests to declare relevant to this article's content.
\end{credits}
%
%
%


%





\appendix 
\section{Appendix: NIS2 Extracted and Marked Clauses} \label{appendix:NIS2clauses}
\begin{adjustwidth}{0cm}{0cm} 
\small 
\setstretch{0.9} 
\sloppy 
\justify 

Art1(1) ``This Directive lays down \textit{measure}s that aim to achieve a high common \textit{level} of cybersecurity across the Union, with a view to improving the functioning of the internal market'' \cite{NIS2}.
 
Art7(2) As part of the national cybersecurity strategy, \underline{Member States} \textbf{shall} in particular \textbf{adopt policies}: (f) promoting and developing education and training on cybersecurity, cybersecurity skills, awareness raising and research and development initiatives, as well as guidance on good cyber hygiene practices and controls, aimed at citizens, stakeholders and \underline{entities}; (i) strengthening the cyber resilience and the cyber hygiene baseline of \underline{small and medium-sized enterprises}, in particular those excluded from the scope of this Directive, by providing easily accessible \textit{guidance and assistance for their specific needs}'' \cite{NIS2}.

Art18(1) requires \underline{ENISA} \textbf{to compile the }\textbf{Report on the state of cybersecurity} in the Union ENISA \textbf{shall adopt}, in cooperation with the Commission and the Cooperation Group, a biennial \textit{report on the state of} \textit{cybersecurity} in the Union and \textbf{shall submit and present} that report to the \underline{European Parliament}. The report \textbf{shall}, inter alia, be made available in machine-readable data and \textbf{include the following}: (b) an \textbf{assessment of}\textbf{ the development of} cybersecurity capabilities in \underline{the public and private sectors} across the Union; (c) \textbf{an assessment of the general level of cybersecurity awareness and cyber hygiene} among citizens and \underline{entities, including small and medium-sized enterprises}; (e) an aggregated \textbf{assessment of the level of maturity} of cybersecurity capabilities and resources across the Union, including those at sector level, as well as of the extent to which the Member States’ national cybersecurity strategies are aligned'' \cite{NIS2}.

Art19(1) The \underline{peer reviews} \textbf{shall cover} at least one of the following: a) the \textbf{level of implementation of the cybersecurity risk-management measures} /.../ 
laid down in Articles 21 
'' \cite{NIS2}.

Art20(1)``\underline{Member States} \textbf{shall} \textbf{ensure} that the \textbf{\underline{management bodies of}}\textbf{\underline{ essential and important entities}} \textit{approve the cybersecurity risk-management measures taken} by those \underline{entities} in order \textit{to comply with Article 21}, \textbf{oversee its implementation} and can be held liable for infringements by the entities of that Article.'' \cite{NIS2}.
    
Art20(2)\underline{Member States} \textbf{shall ensure} that the \underline{members} \underline{of the management} \underline{bodies} \underline{of essential and important entities} are required to follow training, and shall encourage essential and important entities to offer similar training to their \underline{employees} on a regular basis, in order that they gain sufficient knowledge and skills to enable them to identify risks and \textit{assess cybersecurity risk-management practices} and their impact on the services provided by the entity'' \cite{NIS2}.

Art21(1) \underline{Member States} \textbf{shall ensure} that \textbf{\underline{essential and important entities} take appropriate and proportionate technical, operational and organizational measures to manage the risks} posed to the security of network and information systems which those entities use for their operations or for the provision of their services, and to prevent or minimise the impact of incidents on recipients of their services and on other services'' \cite{NIS2}.

Art21(2) The \textit{measure}s referred to in paragraph 1 \textbf{shall be based on an all-hazards approach} that aims to protect \underline{network} and \underline{information systems} and the \underline{physical environment of those systems} from incidents, and shall include at least the following: 
	a) policies on risk analysis and information system security;
	b) incident handling;
	c) business continuity, such as backup management and disaster recovery, and crisis management;
	d) supply chain security, including security-related aspects concerning the relationships between each entity and its direct suppliers or service providers;
	e) security in network and information systems acquisition, development and maintenance, including vulnerability handling and disclosure;
	f) policies and procedures to assess the effectiveness of cybersecurity risk-management measures;
	g) basic cyber hygiene practices and cybersecurity training;
	h) policies and procedures regarding the use of cryptography and, where appropriate, encryption;
	i) human resources security, access control policies and asset management;
	j) the use of multi-factor authentication or continuous authentication solutions, secured voice, video and text communications and secured emergency communication systems within the entity, where appropriate'' \cite{NIS2}.

Art21(3) \underline{Member States} \textbf{shall} \textbf{ensure} that, when considering which measures referred to in paragraph 2, point (d), of this Article are appropriate, \underline{entities} \textit{take into account the vulnerabilities} specific to each direct \underline{supplier and service provider} and the overall quality of products and \textit{cybersecurity practices of their suppliers and service providers}, including their secure development procedures'' \cite{NIS2}.

Art21(4) \underline{Member States} \textbf{shall} \textit{ensure} that \underline{an entity} that finds that it does not comply with the measures provided for in paragraph 2 \textit{takes}, without undue delay, \textit{all necessary, appropriate and proportionate corrective measures}'' \cite{NIS2}.

Art31(2) \underline{Member States}\textbf{ may allow their \underline{competent authorities} to prioritise supervisory tasks}. Such prioritisation s\textbf{hall be based on a risk-based approach}. To that end, when exercising their supervisory tasks provided for in Articles 32 and 33, the \underline{competent authorities} may establish supervisory methodologies allowing for a prioritisation of such tasks following a risk-based approach'' \cite{NIS2}.

Art32(2) \underline{Member States} \textbf{shall ensure} that \underline{the competent authorities}, when exercising their supervisory tasks in relation to \underline{essential entities}, have the power to subject those entities at least to:
a) on-site inspections and \textit{off-site supervision}, including random checks conducted by trained professionals;
e) r\textit{equests for information necessary to \textit{assess} the cybersecurity risk-management measures} adopted by \underline{the entity} concerned, including documented cybersecurity policies, as well as compliance with the obligation to submit information to the \underline{competent authorities} pursuant to Article 27'' \cite{NIS2}.

Art32(4) \underline{Member States} \textbf{shall ensure} that their \underline{competent authorities}, when exercising their enforcement powers in relation to \underline{essential entities}, have the power at least to: d) order the \underline{entities} concerned to \textit{ensure} that their cybersecurity risk-management \textit{measure}s comply with Article 21 /.../  
, in a specified manner and within a specified period;
f) order the \underline{entities} concerned to implement the recommendations provided as a result of a security audit within a reasonable deadline;
g) designate a monitoring officer with well-defined tasks \textbf{for a determined period of time to oversee the compliance} of the \underline{entities} concerned with Articles 21 /.../'' \cite{NIS2}.

Art33(2) \underline{Member States} \textbf{shall ensure} that \underline{the competent authorities}, when exercising their supervisory tasks in relation to \underline{important entities}, have the power to subject those entities at least to: a) \textit{on-site inspections and off-site ex post supervision} conducted by \underline{trained professionals}'' \cite{NIS2}.

Recital (56) \underline{Member States} \textbf{should have} \underline{a point of contact} for \underline{small and medium-sized enterprises} at national or regional level, which either \textit{provides guidance and assistance} to \underline{small and medium-sized enterprises} \textbf{or} \textit{directs them to the appropriate bodies} for guidance and assistance with regard to cybersecurity related issues'' \cite{NIS2}. 

\end{adjustwidth}
\bibliographystyle{splncs04}
\bibliography{mybibliography}

\begin{thebibliography}{10}
\providecommand{\url}[1]{\texttt{#1}}
\providecommand{\urlprefix}{URL }
\providecommand{\doi}[1]{https://doi.org/#1}

\bibitem{alsaadi2019}
Alsaadi, M., Lisitsa, A., Qasaimeh, M.: {Minimizing the ambiguities in medical devices regulations based on software requirement engineering techniques} (2019). \doi{10.1145/3368691.3368709}

\bibitem{bernik2016measuring}
Bernik, I., Prislan, K.: Measuring information security performance with 10 by 10 model for holistic state evaluation  (2016). \doi{https://doi.org/10.1371/journal.pone.0163050}

\bibitem{CHIARA2024105961}
Chiara, P.G.: {Towards a right to cybersecurity in EU law? The challenges ahead}  (2024). \doi{https://doi.org/10.1016/j.clsr.2024.105961}

\bibitem{Cohn2004US}
Cohn, M.: {User Stories Applied: For Agile Software Development} (2004)

\bibitem{Cohn2017US-s}
Cohn, M.: {The Two Ways to Add Detail to User Stories} (2017), \url{https://www.mountaingoatsoftware.com/blog/preview/1691}, last access 2024-04-20

\bibitem{WorldPrivacyForum}
Dixon, P., Emerson, J.: {Global Visualization of Countries with Data Privacy Laws, Treaties, or Conventions}, \url{https://www.worldprivacyforum.org/2024/06/countries-with-data-privacy-laws/}, last access: 2024-11-25

\bibitem{NCSIWeb}
{e-Governance Academy}: National cyber security index. \url{https://ncsi.ega.ee/}, accessed: 2024-04-12

\bibitem{NIS2}
{European Parlament}: {Directive (EU) 2022/2555 of the European Parliament and of the Council of 14 December 2022 on measures for a high common level of cybersecurity across the Union, amending Regulation (EU) No 910/2014 and Directive (EU) 2018/1972, and repealing Directive (EU) 2016/1148 (NIS 2 Directive) } (2022), \url{https://eur-lex.europa.eu/legal-content/en/TXT/?uri=CELEX%3A32022L2555}

\bibitem{ENISAAssessmentSMEs}
{European Union Agency for Cybersecurity}: {Cybersecurity Maturity Assessment for Small and Medium Enterprises}. \url{https://www.enisa.europa.eu/cybersecurity-maturity-assessment-for-small-and-medium-enterprises#/}, accessed: 2024-06-13

\bibitem{EUCSIWeb}
{European Union Agency for Cybersecurity}: {EU Cybersecurity Index}. \url{https://www.enisa.europa.eu/topics/cybersecurity-policy/nis-directive-new/eu-cybersecurity-index}, accessed: 2024-05-20

\bibitem{fatema2016}
Fatema, K., Debruyne, C., Lewis, D., OSullivan, D., Morrison, J.P., Mazed, A.A.: {A Semi-Automated Methodology for Extracting Access Control Rules from the European Data Protection Directive} (2016). \doi{10.1109/SPW.2016.16}

\bibitem{KybermittariWeb}
{Finnsh Transport and Communication Agency National Cyber Security Centre}: Cybermeter. \url{https://www.kyberturvallisuuskeskus.fi/fi/palvelumme/tilannekuva-ja-verkostojohtaminen/kybermittari}, accessed: 2024-05-13

\bibitem{Grigaliūnas2024}
Grigaliūnas, S., Schmidt, M., Brūzgienė, R., Smyrli, P., Andreou, S., Lopata, A.: {Holistic Information Security Management and Compliance Framework}  (2024). \doi{10.3390/electronics13193955}

\bibitem{hassani2024rethinking}
Hassani, S., Sabetzadeh, M., Amyot, D., Liao, J.: {Rethinking Legal Compliance Automation: Opportunities with Large Language Models}  (2024). \doi{10.1109/RE59067.2024.00051}

\bibitem{greeceSelfAssessment}
{Hellenic Ministry of Digital Governance Government department}: {Cybersecurity Self Assessment Tool} (2021), \url{{https://mindigital.gr/wp-content/uploads/2022/03/cybersecurity-self-assessment.xlsm}}, accessed: 2024-04-27

\bibitem{ISO27002:2022}
International Organization for Standardization: ISO/IEC 27002:2022 Information security, cybersecurity and privacy protection — Information security controls (2022)

\bibitem{GCIWeb}
{International Telecommunications Union}: {Global Cybersecurity Index}. \url{https://www.itu.int/en/ITU-D/Cybersecurity/Pages/global-cybersecurity-index.aspx}, accessed: 2024-05-20

\bibitem{Islam2010255}
Islam, S., Mouratidis, H., Wagner, S.: {Towards a framework to elicit and manage security and privacy requirements from laws and regulations}  (2010). \doi{10.1007/978-3-642-14192-8_23}

\bibitem{ISO27001}
{ISO/IEC 27001:2022(en) Information security, cybersecurity and privacy protection — Information security management systems — Requirements}. Standard, International Organization for Standardization (2022)

\bibitem{jazri2018measuring}
Jazri, H., Zakaria, O., Chikohora, E.: Measuring cybersecurity wellness index of critical organisations (2018)

\bibitem{Jorshari2012}
Jorshari, F.Z., Mouratidis, H., Islam, S.: {Extracting security requirements from relevant laws and regulations} (2012). \doi{10.1109/RCIS.2012.6240443}

\bibitem{khaleghi2022comprehensive}
Khaleghi, M., Aref, M.R., Rasti, M.: {Comprehensive Comparison of Security Measurement Models}  (2022). \doi{10.1080/19361610.2021.1981089}

\bibitem{Kiyavitskaya2008}
Kiyavitskaya, N., Krausová, A., Zannone, N.: {Why eliciting and managing legal requirements is hard} (2008). \doi{10.1109/RELAW.2008.10}

\bibitem{leszczyna2021review}
Leszczyna, R.: Review of cybersecurity assessment methods: Applicability perspective  (2021). \doi{https://doi.org/10.1016/j.cose.2021.102376}

\bibitem{Lucassen2016USusability}
Lucassen, G., Dalpiaz, F., Werf, J.M.E.M.v.d., Brinkkemper, S.: {The Use and Effectiveness of User Stories in Practice} (2016). \doi{https://doi.org/10.1007/978-3-319-30282-9_14}

\bibitem{malaivongs2022cybertrust}
Malaivongs, S., Kiattisin, S., Chatjuthamard, P.: Cyber trust index: A framework for rating and improving cybersecurity performance  (2022). \doi{https://doi.org/10.3390/app122111174}

\bibitem{maleh2017towards}
Maleh, Y., Ezzati, A., Sahid, A., Belaissaoui, M.: {Towards A Capability Assessment Framework for Information Security Governance in Organization}  (2017)

\bibitem{CDIWeb}
{MIT Technology Review Insights}: {Cyber Defense Index}. \url{https://www.technologyreview.com/2022/11/15/1063189/the-cyber-defense-index-2022-23/}, accessed: 2024-05-12

\bibitem{EstoniaAuditXroad}
{National Audit Office of Estonia}: Administration and reliability of {X}-road. \url{https://www.riigikontroll.ee/DesktopModules/DigiDetail/FileDownloader.aspx?FileId=14778&AuditId=2520}, accessed: 2024-11-17

\bibitem{EstoniaAuditLocal}
{National Audit Office of Estonia}: Implementation of system of {IT} security measures in local governments. \url{https://www.riigikontroll.ee/DesktopModules/DigiDetail/FileDownloader.aspx?FileId=14270&AuditId=2466}, accessed: 2024-11-17

\bibitem{NUKIBReport}
{National Cyber and Information Security Agency of the Czech Republic}: {2023 Report on the State of Cybersecurity in the Czech Republic}. \url{https://nukib.gov.cz/download/publications_en/2023_Report_on_the_State_of_Cybersecurity_in_the_Czech_Republic.pdf}, accessed: 2023-11-08

\bibitem{pisa2020governing}
Pisa, M., Dixon, P., Ndulu, B., Nwankwo, U.: Governing data for development: trends, challenges, and opportunities  (2020), \url{https://www.cgdev.org/sites/default/files/governing-data-development-trends-challenges-and-opportunities.pdf}

\bibitem{prislan2020real}
Prislan, K., Miheli{\v{c}}, A., Bernik, I.: A real-world information security performance assessment using a multidimensional socio-technical approach  (2020). \doi{https://doi.org/10.1371/journal.pone.0238739}

\bibitem{rae2019defining}
Rae, A., Patel, A.: {Defining a new composite cybersecurity rating scheme for SMEs in the UK} (2019). \doi{https://doi.org/10.1007/978-3-030-34339-2_20}

\bibitem{rea-guaman_maturity_2017}
Rea-Guaman, A.M., Sánchez-García, I.D., Feliu, T.S., Calvo-Manzano, J.A.: Maturity models in cybersecurity: A systematic review (2017). \doi{10.23919/CISTI.2017.7975865}

\bibitem{E-ITS}
{RIA (Estonian Information System Authority}): {E-{ITS}. Portal of {Estonian} Information Security Standard} (2022), \url{https://eits.ria.ee/}

\bibitem{Seeba2022F4SLE}
Seeba, M., M{\"a}ses, S., Matulevi{\v{c}}ius, R.: {Method for Evaluating Information Security Level in Organisations} (2022). \doi{10.1007/978-3-031-05760-1_39}

\bibitem{Seeba2023MUSE}
Seeba, M., amefon Obot~Affia, A., Mäses, S., Matulevičius, R.: {Create your own MUSE: A method for updating security level evaluation instruments}. Computer Standards \& Interfaces  \textbf{87},  103776 (2024). \doi{https://doi.org/10.1016/j.csi.2023.103776}

\bibitem{Seeba2023MASS_F4SLE}
Seeba, M., Oja, T., Murumaa, M.P., Stupka, V.: {Security Level Evaluation with F4SLE}. In: Proceedings of the 18th International Conference on Availability, Reliability and Security. ARES '23, Association for Computing Machinery, New York, NY, USA (2023). \doi{10.1145/3600160.3605045}, \url{https://doi.org/10.1145/3600160.3605045}

\bibitem{szczepaniuk2020information}
Szczepaniuk, E.K., Szczepaniuk, H., Rokicki, T., Klepacki, B.: Information security assessment in public administration  (2020). \doi{https://doi.org/10.1016/j.cose.2019.101709}

\bibitem{tasheva2022}
Tasheva, I., Kunkel, I.: {In a hyperconnected world, is the EU cybersecurity framework connected?}  (2022). \doi{10.1177/17816858221136106}

\bibitem{IASMEEssentials}
{The IASME Consortium}: {IASME Cyber Essentials}. \url{https://getreadyforcyberessentials.iasme.co.uk/}, accessed: 2024-06-13

\bibitem{irelandSelfAssessment}
{The National Cyber Security Centre of Ireland}: {Cyber Security Baseline Standards Self-Assessment Form} (2023), \url{{https://www.ncsc.gov.ie/pdfs/Cyber_Resilience_Self-Assessment_Framework_Version_1.4_Jan_23.xlsx}}, accessed: 2024-05-27

\bibitem{spainSelfAssessment}
{The Spanish National Cybersecurity Institute}: {Herramienta de Autodiagnóstico }, \url{{https://adl.incibe.es/#}}, accessed: 2024-05-27

\bibitem{latviaAudit2022eservices}
{The State Audit Office of the Republic of Latvia}: {Can we rely on the access to information systems and the receipt of e-services?} \url{https://www.lrvk.gov.lv/en/getrevisionfile/29525-5Aio6j7MwYsuSG4nKlzFVmCMG0JZircA.pdf} (2022)

\bibitem{C2M2Tool}
{United States Department of Energy}: {C2M2 C2M2 V2.1 HTML-Based Tool}. \url{https://c2m2.doe.gov/c2m2-assessment}, accessed: 2024-06-13

\bibitem{BelferCent2020NCPI}
{University of Harvard Belfer Center}: National cyber power index 2020. \url{https://www.belfercenter.org/publication/national-cyber-power-index-2022} (2022)

\bibitem{VANDEZANDE2024105890}
Vandezande, N.: {Cybersecurity in the EU: How the NIS2-directive stacks up against its predecessor}  (2024). \doi{https://doi.org/10.1016/j.clsr.2023.105890}

\bibitem{Wanecki2023}
Wanecki, P., Jasek, R., Drofova, I.: {The Contribution of the European NIS2 Directive to the Design of the Cyber Security Model} (2023). \doi{10.1109/IDT59031.2023.10194454}

\bibitem{you2016advanced}
You, Y., Cho, I., Lee, K.: An advanced approach to security measurement system  (2016). \doi{https://doi.org/10.1007/s11227-015-1585-7}

\bibitem{yu2010social}
Yu, E., Giorgini, P., Maiden, N., Mylopoulos, J.: {Social modeling for requirements engineering: An introduction}  (2010). \doi{10.7551/mitpress/7549.001.0001}

\end{thebibliography}
\end{document}